\documentclass[RNAAS]{aastex62}

\newcommand{\kms}{km\,s$^{-1}$}
\newcommand{\Mnom}{\hbox{$\mathcal{M}^{\rm N}_{\odot}$}}

\newcommand{\Rnom}{\hbox{$\mathcal{R}^{\rm N}_{\odot}$}}

\begin{document}

\shorttitle{BW Aqr} 
\title{Absolute parameters for the F-type eclipsing binary BW Aquarii}

\author[0000-0003-3794-1317]{P. F. L. Maxted} 
\affiliation{Astrophysics Group, Keele University, Staffordshire, ST5 5BG, UK}
\email{p.maxted@keele.ac.uk}


\maketitle

 BW Aqr is a bright eclipsing binary star containing a pair of F7V stars. The
absolute parameters of this binary (masses, radii, etc.) are known to good
precision so they are often used to test stellar models, particularly in
studies of convective overshooting \citep[e.g., ][]{1991A&A...246..397C,
2018arXiv180403148C}.

% BW Aqr was observed as part of Kepler K2 Campaign 3
%\citet{2014PASP..126..398H}. \citet{2018arXiv180310522M} analysed this light
\citet{2018arXiv180310522M} analysed the Kepler K2 data
for BW~Aqr and noted that it shows variability
between the eclipses that may be caused by tidally induced pulsations. The
authors warn that they had ``not attempted to characterise the level of
systematic error''  in the parameters they derived.
\citet{2018arXiv180502670L} analysed the same data
together with new radial velocity (RV) measurements. The binary star model
they used gave a poor fit to the light curve through the secondary eclipse so
the radii they derive may be subject to significant systematic error. The
standard errors they quote on the stellar masses based on $N=14$ pairs of RV
measurements are clearly underestimated. The RMS (root-mean square) of the
residuals from their spectroscopic orbit fit to these RVs is $\sigma \approx
3.6$\,\kms\ so the semi-amplitudes of the spectroscopic orbit are expected to
have errors $\sigma_K\approx \sigma/\sqrt{N/2} = 1.4$\,\kms\
\citep{1999DSSN...13...28M}, cf. the quoted errors of 0.27\,\kms.

Table 1 shows the absolute parameters for BW Aqr derived from an improved
analysis of the Kepler K2 light curve plus the RV measurements from both
\citet{1979A&AS...36..453I} and \citet{2018arXiv180502670L}. The light curve
data used are identical to those shown in \citet{2018arXiv180310522M}. We used
{\sc ellc} version 1.8.0\footnote{\url{pypi.org/project/ellc/} } to model the light curve
using the power-2 limb darkening law, $I_{\lambda}(\mu) = 1 - c\left(
  1-\mu^{\alpha}\right)$, with the same parameters $c$ and $\alpha$ for both
stars and with Gaussian priors on the
parameters $h_1 =1-c\left(1-2^{-\alpha}\right) = 0.78\pm 0.02$ and $h_2 =
c2^{-\alpha} = 0.44 \pm 0.10$ \citep{2018arXiv180407943M}. The ``default''
grid size in {\sc ellc} was used so that numerical noise is less than 60\,ppm.
Other details of the fit to the light curve are similar to those described in
\citet{2018arXiv180310522M}.
%The exposure time of 1798s is accounted for using numerical integration over 5
%model light curve points per observation.

Each pair of consecutive primary and secondary eclipses in the K2 light curve
were analysed separately using the {\sc emcee} algorithm to determine the
median of the posterior probability distributions (PPDs) for the model
parameters. The results shown in Table 1 are calculated from the mean and its
standard error from these 10 median values. The RV measurements were fit
simultaneously with the times of mid-eclipse from \citet{2018arXiv180502670L}
and \citet{2014CoSka..43..419V} using the model {\sc omdot} to account for
apsidal motion \citep{2015A&A...578A..25M}. 
%Standard errors were assigned to the observations used in {\sc omdot} such
%that the contribution of each data set to the $\chi^2$ value of the best fit
%is approximately equal to the number of observations in that data set and the
%reduced $\chi^2$ value of the best fit is close to 1. Absolute parameters
%were derived using {\sc jktabsdim} version ??? with IAU recommended nominal
%values for solar constants \cite{2016AJ....152...41P}. 
The values in Table 1 with their robust error estimates from the standard
deviation of the mean are consistent with the values and errors from
\citet{2018arXiv180310522M} based on the PPD calculated using {\sc emcee} for
a fit to the entire K2 light curve.

{\catcode`\&=11 \gdef\1991C{\citep{1991A&A...246..397C}}}
{\catcode`\&=11 \gdef\2010C{\citep{2010A&A...516A..42C}}}

\begin{deluxetable*}{lcRRl}
\tablecaption{Model parameters and absolute parameters for BW Aquarii.
}
\tablehead{\colhead{Parameter} & \colhead{Units} & \colhead{Value} &
\colhead{Standard error} & \colhead{Notes} }
%% All data must appear between the \startdata and \enddata commands
\startdata
  \multicolumn{4}{l}{Model parameters, {\sc ellc}} \\
$P_{\rm s}$      &  d         &  6.719695     &            & Siderial period (fixed value for light curve analysis only) \\
$(R_1+R_2)/a$    &            &  0.17916      &   0.00024  & Sum of stellar radii relative to orbital semi-major axis \\
$R_2/R_1$        &            &  1.1948       &   0.0049   & \\
$S_{\rm Kp}$     &            &  0.9291       &   0.0014   & Surface brightness ratio in the Kepler bandpass \\
$i$              & $^{\circ}$ &  88.65        &  0.03      & Orbital inclination \\
$f_c$            &            & -0.09415      &  0.00014   & $\sqrt{e}\cos \omega $\\
$f_s$            &            &  0.41081      &  0.00047   & $\sqrt{e}\sin \omega $\\
$h_1$            &            &  0.7974       &  0.0066    &  \\
$h_2$            &            &  0.251        &  0.070     &  \\
$y$              &            &  0.1          &  0.2       & Monochromatic gravity darkening exponent \\
$\sigma_f$       & \%         &  0.017      &  0.005   & Standard error of K2 flux measurements \\
  \multicolumn{4}{l}{Quantities derived from {\sc ellc} model parameters} \\
$e$              &            &  0.17763      &  0.00036   & Orbital eccentricity  \\
$\omega$         & $^{\circ}$ &  102.91       &  0.03      & Longitude of periastron  \\
$R_1/a $         &            &  0.08163      &  0.00028   & Radius of star 1 relative to orbital semi-major axis \\
$R_2/a $         &            &  0.09753      &  0.00010   & Radius of star 2 relative to orbital semi-major axis \\
$\ell_{\rm Kp}$  &            &  1.326        &  0.011     & Flux ratio in the Kepler bandpass \\ 
\multicolumn{4}{l}{Model parameters, {\sc omdot}} \\
T$_0$            &            & 2456990.46693 & 0.00008    & Reference time of mid-eclipse, BJD (TDB)\\
P$_{\rm a}$      &  d         &    6.71971325 & 0.00000057 & Anomalistic orbital period \\
e                &            &       0.1771  & 0.0042     & Orbital eccentricity  \\
$\omega_0$       & $^{\circ}$ &         103.0 & 0.3        & Longitude of periastron at time T$_0$  \\
$\dot{\omega}$   &  d$^{-1}$  &     0.0000197 & 0.0000006  & Apsidal motion rate, radians per siderial period \\
$\gamma$         &  \kms      &          9.73 & 0.25       & Systemic radial velocity \\
K$_1$            & \kms       &         83.93 &       0.45 & Semi-amplitude of spectroscopic orbit, star 1 \\
K$_2$            & \kms       &         78.42 &       0.47 & Semi-amplitude of spectroscopic orbit, star 2 \\
$\sigma_{RV, 1}$ & \kms       &  3.12         &            & RMS of residuals for all radial velocity data, star 1 \\
$\sigma_{RV, 2}$ & \kms       &  2.78         &            & RMS of residuals for all radial velocity data, star 2 \\
$\sigma_{t}    $ & d          &  0.00055      &            & RMS of residuals for times of mid-eclipse \\ 
\multicolumn{4}{l}{Quantities derived from {\sc omdot} model parameters} \\
U                & y          & 5880          & 180        & Apsidal motion period \\
\multicolumn{4}{l}{Absolute parameters} \\
T$_{\rm eff, 1}$ & K          &  6450         & 100        & Effective temperature, star 1 \1991C  \\
T$_{\rm eff, 2}$ & K          &  6350         & 100        & Effective temperature, star 2 \1991C \\
%E$(\rm B-\rm V)$ &            &  0\fm04       & 0\fm02     & Assuming 50\% error in  ``$A_{\rm V}\approx0\fm 13$'' from \1991C \\
%$[{\rm Fe/H}]$   & dex        & $-0.07$       & 0.11       & \2010C \\  
$M_1$            & \Mnom      &   1.373       & 0.018      & Mass, star 1 \\
$M_2$            & \Mnom      &   1.469       & 0.018      & Mass, star 2 \\
$R_1$            & \Rnom      &   1.7332      & 0.0091     & Radius, star 1 \\
$R_2$            & \Rnom      &   2.0708      & 0.0086     & Radius, star 2 \\
$\log g_1$       &cm\,s$^{-2}$&   4.0982      & 0.0040     & Surface gravity, star 1 \\
$\log g_2$       &cm\,s$^{-2}$&   3.9731      & 0.0025     & Surface gravity, star 2 \\
$\log{\rm L}_1$   & L$_{\sun}$ &    0.669     &   0.027     & Luminosity, star 1 \\
$\log{\rm L}_2$   & L$_{\sun}$ &    0.796     &   0.028     & Luminosity, star 2 \\
\enddata
\end{deluxetable*}

\acknowledgments
PM gratefully acknowledges support provided by the UK Science and Technology
Facilities Council through grant number ST/M001040/1.

\facility{Kepler}
\software{{\sc ellc} \citep{2016A&A...591A.111M}, 
{\sc emcee} \citep{2013PASP..125..306F}}

\bibliography{mybib}{}
\bibliographystyle{aasjournal}
\end{document}